\documentclass{article}
\usepackage{amsmath}

\input{tcilatex}

\begin{document}

\author{\textbf{Howard E. Brandt} \\
U.S. Army Research Laboratory, Adelphi, MD\\
hbrandt@arl.army.mil \and \textbf{John M. Myers} \\
Gordon McKay Laboratory, Harvard University, Cambridge, MA\\
myers@deas.harvard.edu}
\title{\textbf{Expanded Quantum Cryptographic Entangling Probe}}
\maketitle

\begin{abstract}
The paper [Howard E. Brandt, "Quantum Cryptographic Entangling Probe," Phys.
Rev. A \textbf{71}, 042312 (2005)] is generalized to include the full range
of error rates for the projectively measured quantum cryptographic
entangling probe.


\end{abstract}

\section{INTRODUCTION}

Recently, a design was presented \cite{PRA-05}, \cite{ModOp-05} for an
optimized entangling probe attacking the BB84 Protocol \cite{Bennett1} of
quantum key distribution (QKD) and yielding maximum Renyi information to the
probe for a set error rate induced by the probe. Probe photon polarization
states become optimally entangled with the BB84 signal states on their way
between the legitimate transmitter and receiver. Standard von Neumann
projective measurements of the probe yield maximum information on the
pre-privacy amplified key, once basis information is revealed during
reconciliation. A simple quantum circuit was found, consisting of a single
CNOT gate, and faithfully producing the optimal entanglement. The control
qubit consists of two photon polarization-basis states of the signal, the
target qubit consists of two probe photon polarization basis states, and the
initial state of the probe is set by an explicit algebraic function of the
error rate to be induced by the probe. A method was determined for measuring
the appropriate probe states correlated with the BB84 signal states and
yielding maximum Renyi information to the probe. It was assumed throughout
that the error rate $E$ induced by the probe in the legitimate signal was
such that $0\leq E\leq 1/4$ for the projectively measured probe. Here we
extend the analysis to cover the full range of theoretical interest, namely $%
0\leq E\leq 1/3.$

\section{GENERALIZED ENTANGLING PROBE}

In the present work a generalization is given to include the full range of
error rates, $0\leq E\leq 1/3$. To accomplish this, the following sign
choices must be made for the probe parameter $\mu $ in Eqs. (26) and (27) of 
\cite{PRA-05}:%
\begin{equation}
\ \cos \mu =[(1+\eta )/2]^{1/2},\ \ \ \ 
\end{equation}%
\begin{equation}
\ \sin \mu =\text{sgn}(1-4E)[(1-\eta )/2]^{1/2},\ \ \ \ 
\end{equation}%
in which we define%
\begin{equation}
\ \text{sgn}(x)\equiv \left\{ 
\begin{array}{c}
1,\ \ \ x>0\ \ \ \ \  \\ 
\ \ 0,\ \ \ x=0\ \ \ \ \ \ \  \\ 
-1,\ \ \ x<0\ \ \ \ \ \ \ \ 
\end{array}%
\right. \ 
\end{equation}%
One also has the definition, Eq. (75) of \cite{PRA-05}:%
\begin{equation}
\eta \equiv \left[ 8E(1-2E)\right] ^{1/2}.
\end{equation}%
In this case, the probe states $\left\vert A_{1}\right\rangle $, $\left\vert
A_{2}\right\rangle $, $\left\vert \alpha _{+}\right\rangle $, $\left\vert
\alpha _{-}\right\rangle $, and $\left\vert \alpha \right\rangle $ of \cite%
{PRA-05} become:%
\begin{equation}
\left\vert A_{1}\right\rangle \equiv \left[ \frac{1}{2}(1+\eta )\right]
^{1/2}\left\vert w_{0}\right\rangle +\ \text{sgn}(1-4E)\left[ \frac{1}{2}%
(1-\eta )\right] ^{1/2}\left\vert w_{3}\right\rangle ,
\end{equation}%
\begin{equation}
\left\vert A_{2}\right\rangle \equiv \ \text{sgn}(1-4E)\left[ \frac{1}{2}%
(1-\eta )\right] ^{1/2}\left\vert w_{0}\right\rangle +\left[ \frac{1}{2}%
(1+\eta )\right] ^{1/2}\left\vert w_{3}\right\rangle ,
\end{equation}%
\begin{eqnarray}
\left\vert \alpha _{+}\right\rangle  &=&\left[ \left( 2^{1/2}+1\right)
\left( 1+\eta \right) ^{1/2}+\text{sgn}(1-4E)\left( 2^{1/2}-1\right) \left(
1-\eta \right) ^{1/2}\right] \left\vert w_{0}\right\rangle   \notag \\
&&+\left[ \text{sgn}(1-4E)\left( 2^{1/2}+1\right) \left( 1-\eta \right)
^{1/2}+\left( 2^{1/2}-1\right) \left( 1+\eta \right) ^{1/2}\right]
\left\vert w_{3}\right\rangle ,\ \ \ \ \ \ \ \ \ \ \ \ \ 
\end{eqnarray}%
\begin{eqnarray}
\left\vert \alpha _{-}\right\rangle  &=&\left[ \left( 2^{1/2}-1\right)
\left( 1+\eta \right) ^{1/2}+\text{sgn}(1-4E)\left( 2^{1/2}+1\right) \left(
1-\eta \right) ^{1/2}\right] \left\vert w_{0}\right\rangle   \notag \\
&&+\left[ \text{sgn}(1-4E)\left( 2^{1/2}-1\right) \left( 1-\eta \right)
^{1/2}+\left( 2^{1/2}+1\right) \left( 1+\eta \right) ^{1/2}\right]
\left\vert w_{3}\right\rangle ,\ \ \ \ \ \ \ \ \ \ \ \ \ 
\end{eqnarray}%
\begin{eqnarray}
\left\vert \alpha \right\rangle  &=&\left[ \text{sgn}(1-4E)\left( 1-\eta
\right) ^{1/2}-\left( 1+\eta \right) ^{1/2}\right] \left\vert
w_{0}\right\rangle   \notag \\
&&+\left[ \left( 1+\eta \right) ^{1/2}-\text{sgn}(1-4E)\left( 1-\eta \right)
^{1/2}\right] \left\vert w_{3}\right\rangle ,\ \ \ \ \ \ \ \ \ \ \ \ 
\end{eqnarray}%
\newline
respectively, where $\left\vert w_{0}\right\rangle $ and $\left\vert
w_{3}\right\rangle $ are the orthonormal basis states in the two-dimensional
Hilbert space of the probe. As in \cite{PRA-05}, the upper sign choice in
Eq. (23) of \cite{PRA-05} has been chosen. Note that Eqs. (5)-(9) are
consistent with Eqs. (207), (210), (204), (205), and (74) of \cite{PRA-05}
for $0\leq E\leq 1/4$, as must be the case.$\ \ $It then follows that Eq.
(71) of \cite{PRA-05}, along with Eqs. (7)-(9) above, now apply for $0\leq
E\leq 1/3$. (Note that $E=1/3\ $corresponds to complete information gain by
the quantum cryptographic entangling probe.) Also the probe and measurement
implementations remain the same (as in \cite{PRA-05}, \cite{ModOp-05}) with
the initial state of the probe now given by Eq. (6). In obtaining the
maximum Renyi information gain $I_{opt}^{R}$ by the probe, Eq. (208) of \cite%
{PRA-05}, from Eqs. (7) and (8) above and Eqs. (23) and (17) of \cite{HB-1}
and the discussion following Eq. (75) of \cite{PRA-05}, one first has 
\begin{equation}
I_{opt}^{R}=\log _{2}(2-Q^{2}),
\end{equation}%
and one readily obtains for the overlap $Q$ of correlated probe states: 
\begin{equation}
Q=\frac{\left\langle \alpha _{+}|\alpha _{-}\right\rangle }{|\alpha
_{+}||\alpha _{-}|}=\frac{1+3\text{sgn}(1-4E)(1-\eta ^{2})^{1/2}}{3+\text{sgn%
}(1-4E)(1-\eta ^{2})^{1/2}}.
\end{equation}%
Then substituting Eq.\ (4) in Eq. (11), one obtains 
\begin{equation}
Q=\frac{1+3\text{sgn}(1-4E)((1-4E)^{2})^{1/2}}{3+\text{sgn}%
(1-4E)((1-4E)^{2})^{1/2}},
\end{equation}%
where we mean the positive square root; i.e. 
\begin{equation}
((1-4E)^{2})^{1/2}=|1-4E|.
\end{equation}%
On noting that 
\begin{equation}
\text{sgn}(1-4E)|1-4E|=1-4E,
\end{equation}%
and substituting Eqs. (13) and (14) in Eq. (12), one obtains 
\begin{equation}
Q=\frac{1-3E}{1-E}.
\end{equation}%
Finally, substituting Eq. (15) in Eq. (10), one obtains Eq. (208) of \cite%
{PRA-05} , namely, 
\begin{equation}
I_{opt}^{R}=\log _{2}\left[ 2-\left( \frac{1-3E}{1-E}\right) ^{2}\right] ,
\end{equation}%
for the full range of error rates, $0\leq E\leq 1/3$, as required.

\section{CONCLUSION}

The quantum cryptographic entangling probe defined in \cite{PRA-05}, \cite%
{ModOp-05} has been generalized to include the full range of error rates, $%
0\leq E\leq 1/3,$ induced by the probe.

\section{ACKNOWLEDGEMENTS}

This work was supported by the U.S. Army Research Laboratory and the Defense
Advanced Research Projects Agency.

\bigskip

\end{document}